\documentclass[%
reprint,
superscriptaddress,
showpacs,preprintnumbers,
amsmath,amssymb,
prb,
]{revtex4-1}
\usepackage{graphicx}
\usepackage{dcolumn}
\usepackage{xfrac}
\usepackage{bm}
\usepackage{color}
\usepackage[export]{adjustbox}
\usepackage{xr-hyper}
\usepackage{hyperref}
\usepackage{tabularx}

\begin{document}

\title{Topological spiral magnetism in the Weyl semimetal SmAlSi}

\author{Xiaohan~Yao}
\affiliation{Department of Physics, Boston College, Chestnut Hill, MA 02467, USA}

\author{Jonathan~Gaudet}
\affiliation{NIST Center for Neutron Research, Gaithersburg, Maryland 20899, USA}
\affiliation{Department of Materials Science and Eng., University of Maryland, College Park, MD 20742-2115}

\author{Rahul~Verma}
\affiliation{Department of Condensed Matter Physics and Materials Science, Tata Institute of Fundamental Research, Colaba, Mumbai 400005, India}

\author{David~E.~Graf}
\affiliation{National High Magnetic Field Laboratory, Tallahassee, FL 32310, USA}

\author{Hung-Yu~Yang}
\affiliation{Department of Physics, Boston College, Chestnut Hill, MA 02467, USA}

\author{Faranak~Bahrami}
\affiliation{Department of Physics, Boston College, Chestnut Hill, MA 02467, USA}

\author{Ruiqi~Zhang}
\affiliation{Department of Physics, Tulane University, New Orleans, LA 70118, USA}

\author{Adam~A.~Aczel}
\affiliation{Neutron Scattering Division, Oak Ridge National Laboratory, Oak Ridge, Tennessee 37831, USA}

\author{Sujan~Subedi}
\affiliation{Department of Physics, Temple University, Philadelphia, PA 19122, USA}

\author{Darius~H.~Torchinsky}
\affiliation{Department of Physics, Temple University, Philadelphia, PA 19122, USA}

\author{Jianwei~Sun}
\affiliation{Department of Physics, Tulane University, New Orleans, LA 70118, USA}

\author{Arun~Bansil}
\affiliation{Department of Physics, Northeastern University, Boston, MA 02115, USA}

\author{Shin-Ming~Huang}
\affiliation{Department of Physics, National Sun Yat-sen University, Kaohsiung 80424, Taiwan}

\author{Bahadur~Singh}
\affiliation{Department of Condensed Matter Physics and Materials Science, Tata Institute of Fundamental Research, Colaba, Mumbai 400005, India}

\author{Predrag~Nikoli\'{c}}
\affiliation{Department of Physics and Astronomy, George Mason University, Fairfax, Virginia 22030, USA}
\affiliation{Institute for Quantum Matter at Johns Hopkins University, Baltimore, MD 21218, USA}

\author{Peter~Blaha}
\affiliation{Institute of Materials Chemistry, Vienna University of Technology, 1060 Vienna, Austria}

\author{Fazel~Tafti}
\email{fazel.tafti@bc.edu}
\affiliation{Department of Physics, Boston College, Chestnut Hill, MA 02467, USA}


\begin{abstract}
Weyl electrons are intensely studied due to novel charge transport phenomena such as chiral anomaly, Fermi arcs, and photogalvanic effect.
Recent theoretical works suggest that Weyl electrons can also participate in magnetic interactions, and the Weyl-mediated indirect exchange coupling between local moments is proposed as a new mechanism of spiral magnetism that involves chiral electrons.
Despite reports of incommensurate and non-collinear magnetic ordering in Weyl semimetals, an actual spiral order has remained hitherto undetected. 
Here, we present evidence of Weyl-mediated spiral magnetism in SmAlSi from neutron diffraction, transport, and thermodynamic data.
We show that the spiral order in SmAlSi results from the nesting between topologically non-trivial Fermi pockets and weak magnetocrystalline anisotropy, unlike related materials (Ce,Pr,Nd)AlSi, where a strong anisotropy prevents the spins from freely rotating.
We map the magnetic phase diagram of SmAlSi and reveal an A-phase where topological magnetic excitations may exist.
This is corroborated by the observation of a topological Hall effect within the A-phase.
\end{abstract}

\maketitle


\section{\label{sec:introduction}Introduction}
\begin{figure}
	\includegraphics[width=0.48\textwidth]{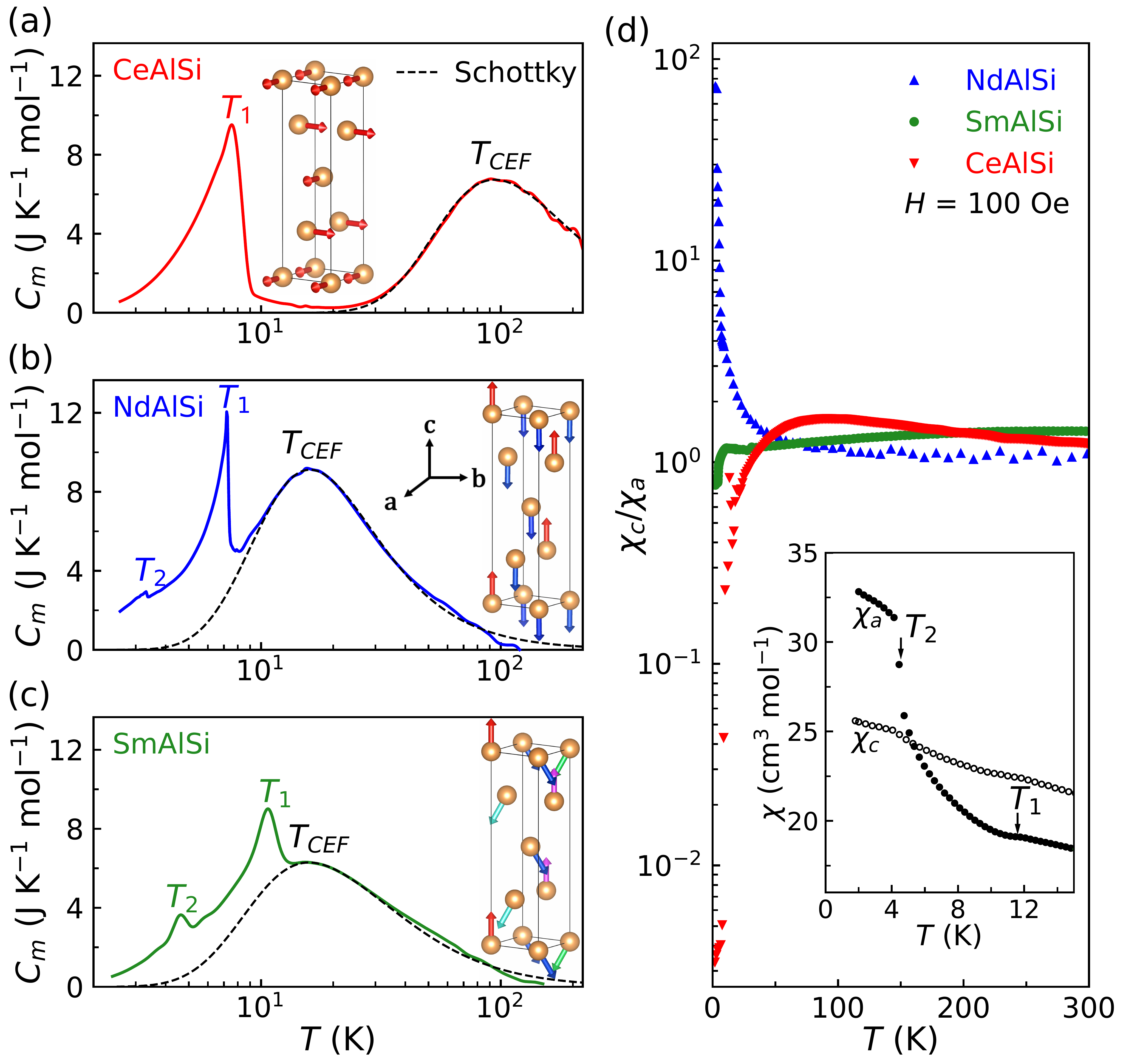}
	\caption{\label{fig:CEF}
	Magnetic heat capacity is plotted as a function of temperature in (a) CeAlSi, (b) NdAlSi, and (c) SmAlSi.
	The broad Schottky anomaly is due to the crystal field levels and the sharp peaks are magnetic transitions.
	The insets show an in-plane anisotropy in CeAlSi, out-of-plane anisotropy in NdAlSi, and spiral order in SmAlSi, consistent with the $\chi_c/\chi_a$ ratios plotted in (d).
	Whereas CeAlSi and NdAlSi show strong in-plane and out-of-plane anisotropy, SmAlSi shows $\chi_c/\chi_a$ of order one.
	}
\end{figure}
It is well known that the combination of magnetism and topology leads to novel electromagnetic phenomena such as extreme magnetoresistance, quantum anomalous Hall effect, axion insulator, and chiral domain walls~\cite{tokura_magnetic_2019,wang_colossal_2021,bestwick_precise_2015,hu_realization_nodate,liu_robust_2020,sun_mapping_2021,xu_back_2021}.
It is less known, however, whether topological entities such as Dirac and Weyl fermions can directly participate in magnetic interactions and drive specific types of magnetic ordering.
Recent calculations have shown that chiral Weyl electrons could couple to local magnetic moments through the Ruderman-Kittel-Kasuya-Yosida (RKKY) mechanism, and mediate indirect Heisenberg ($J_{ij}\textbf{S}_i\cdot\textbf{S}_j$), Kitaev ($K_\gamma S^{\gamma}_i S^{\gamma}_j$), and Dzyaloshinskii-Moriya (DM) interactions ($\textbf{D}\cdot \textbf{S}_i\times\textbf{S}_j$)~\cite{nikolic_universal_2021,nikolic_dynamics_2021,wang_rkky_2017,chang_rkky_2015}.
The Weyl-mediated RKKY interactions provide a new mechanism for spiral magnetic ordering and topological magnetic defects (hedghogs, skyrmions, and meron-antimeron pairs) that are in demand for high-density and high-speed memory devices~\cite{wang_meron_2021,repicky_atomic-scale_2021,luo_skyrmion_2021}.

Experimental evidence of such a Weyl-mediated RKKY interaction has recently been found in NdAlSi~\cite{gaudet_weyl-mediated_2021}, a Weyl semimetal with an incommensurate modulated spin density wave (SDW) whose wavelength is linked to the nesting vector between two topologically non-trivial Fermi pockets. 
A chiral transverse component is observed for this SDW, which is promoted by the Weyl-mediated DM interactions. 
However, a strong magnetocrystalline anisotropy (MCA) in NdAlSi prevents the spins from developing a truly spiral order. Instead, a commensurate ferrimagnetic order is ultimately stabilized at low temperatures~\cite{gaudet_weyl-mediated_2021}.

In this letter, we study SmAlSi, a true spiral magnetic counterpart of NdAlSi due to a much weaker MCA. 
We use heat capacity, neutron diffraction, magnetization, Hall effect, band structure calculations and Monte Carlo simulations to reveal a spiral magnetic order in SmAlSi at low temperatures and attribute it to the Weyl-induced spin interactions in the absence of a significant MCA. We also reveal a region in the phase diagram of SmAlSi that resembles the A-phase in helimagnets such as MnSi and Gd$_3$Ru$_4$Al$_{12}$, where skyrmions have been found~\cite{neubauer_topological_2009,hirschberger_skyrmion_2019}.

\section{\label{sec:methods}Methods}
Single crystals of SmAlSi~were grown using a self-flux method~\cite{xu_shubnikov-haas_2021}.
A Bruker D8 ECO system was used for powder X-ray diffraction, and the FullProf suite was used for the Rietveld refinement~\cite{rodriguez-carvajal_recent_1993}.
The electrical transport, heat capacity, and magnetic susceptibility were measured using the Quantum Design PPMS Dynacool and MPMS-3 instruments. 
A piezoresistive technique was used for the high-field magnetometry at the National High Magnetic Field Laboratory.    
The neutron diffraction experiment was performed with the fixed-incident-energy (14.5~meV) neutron triple-axis spectrometer HB-1A in the High Flux Isotope Reactor (HFIR) at the Oak Ridge National Laboratory, and the SARAh software was used for the representational analysis of the magnetic structure~\cite{SARAh}.
Density functional theory (DFT) calculations with the full-potential linearized augmented plane-wave (LAPW) method were implemented in the WIEN2k code~\cite{blaha_wien2k_2020} using the Perdew-Burke-Ernzerhof (PBE) exchange-correlation functional~\cite{perdew_generalized_1996}, spin-orbit coupling (SOC), and on-site Coulomb repulsion (Hubbard $U$)~\cite{anisimov_band_1991}.
Monte Carlo simulations were performed using the Metropolis algorithm.

\section{\label{sec:results}Results and Discussion}
\begin{table}
	\caption{\label{tab:TCEF}
	$T_{\text{CEF}}$ and $\Delta$ are extracted from the Schottky fits in Fig.~\ref{fig:CEF} (see Fig.~S2).
	The softening of MCA is reflected in the systematic reduction of $T_{\text{CEF}}$, $\Delta$, and the $\chi_c/\chi_a$ ratio at 2~K.
	The PrAlSi data are reproduced from Ref.~\cite{lyu_nonsaturating_2020}.}
	\begin{ruledtabular}
		\begin{tabular}{llllll}			
			Material & $T_{\text{CEF}}$~(K) & $\Delta$~(meV) & $T_1$~(K) & $T_2$~(K) & $\chi_c/\chi_a$ \\
			\hline
			CeAlSi & 91(7) & 29(11) & 8.2(2)  & N/A & (188)$^{-1}$  \\
			PrAlSi & 30(6) & 18(9) & 17.8(2) & N/A & 194           \\
			NdAlSi & 16(3) & 6(2)  & 7.2(1)  & 3.3(1) & 84            \\
			SmAlSi & 15(3) & 5(2)  & 10.6(2) & 4.6(2) & 0.8           \\
		\end{tabular}
	\end{ruledtabular}
\end{table}
\emph{Heat Capacity.} SmAlSi belongs to a family of RAlSi Weyl semimetals that have magnetic rare-earth (R$^{3+}$) ions in a non-centrosymmetric space group $I4_1md$ (inset of Fig.~\ref{fig:CEF} and Fig.~S1)~\cite{suppmatt,chang_magnetic_2018,lyu_nonsaturating_2020,yang_noncollinear_2021}.
The key parameter that enables spiral ordering in SmAlSi, unlike in other RAlSi compounds, is a weak MCA that allows the spins to rotate nearly freely.
The MCA can be quantified using the heat capacity and magnetic susceptibility data.
We isolate the magnetic heat capacity ($C_m$) of RAlSi in Fig.~\ref{fig:CEF} by subtracting the phonon background using the non-magnetic isostructural compound LaAlSi (Fig.~S2).
The broad Schottky anomaly at $T_{\text{CEF}}$ in Figs.~\ref{fig:CEF}a,b,c is due to crystal electric field (CEF) excitations, and the sharp peaks at $T_1$ and $T_2$ are due to magnetic ordering.
We model the Schottky anomaly by assuming a doublet ground-state separated by excited multiplets in the Kramers ions Ce$^{3+}$, Nd$^{3+}$, and Sm$^{3+}$ (Fig.~S2)~\cite{suppmatt}. 
Our Schottky fits (black dashed lines in Figs.~\ref{fig:CEF}a,b,c) yield CEF gaps $\Delta=29(11)$, $6(2)$, and $5(2)$~meV in CeAlSi, NdAlSi, and SmAlSi, respectively. 
The systematic reduction of $\Delta$ and $T_{\text{CEF}}$ in Table~\ref{tab:TCEF} shows a general softening of the MCA from R = Ce to Nd and Sm.
This is also reflected in the systematic change of the ratio between out-of-plane and in-plane magnetic susceptibility $\chi_c/\chi_a$ at 2~K (Fig.~\ref{fig:CEF}d and Table~\ref{tab:TCEF}).
Indeed, whereas CeAlSi has an easy plane with $\chi_a/\chi_c\approx 200$, and NdAlSi has an easy axis with $\chi_c/\chi_a\approx 80$, SmAlSi has a negligible MCA with $\chi_c/\chi_a\approx 1$ (inset of Fig.~\ref{fig:CEF}d), hence its propensity toward spiral ordering.

\begin{figure*}
	\includegraphics[width=\textwidth]{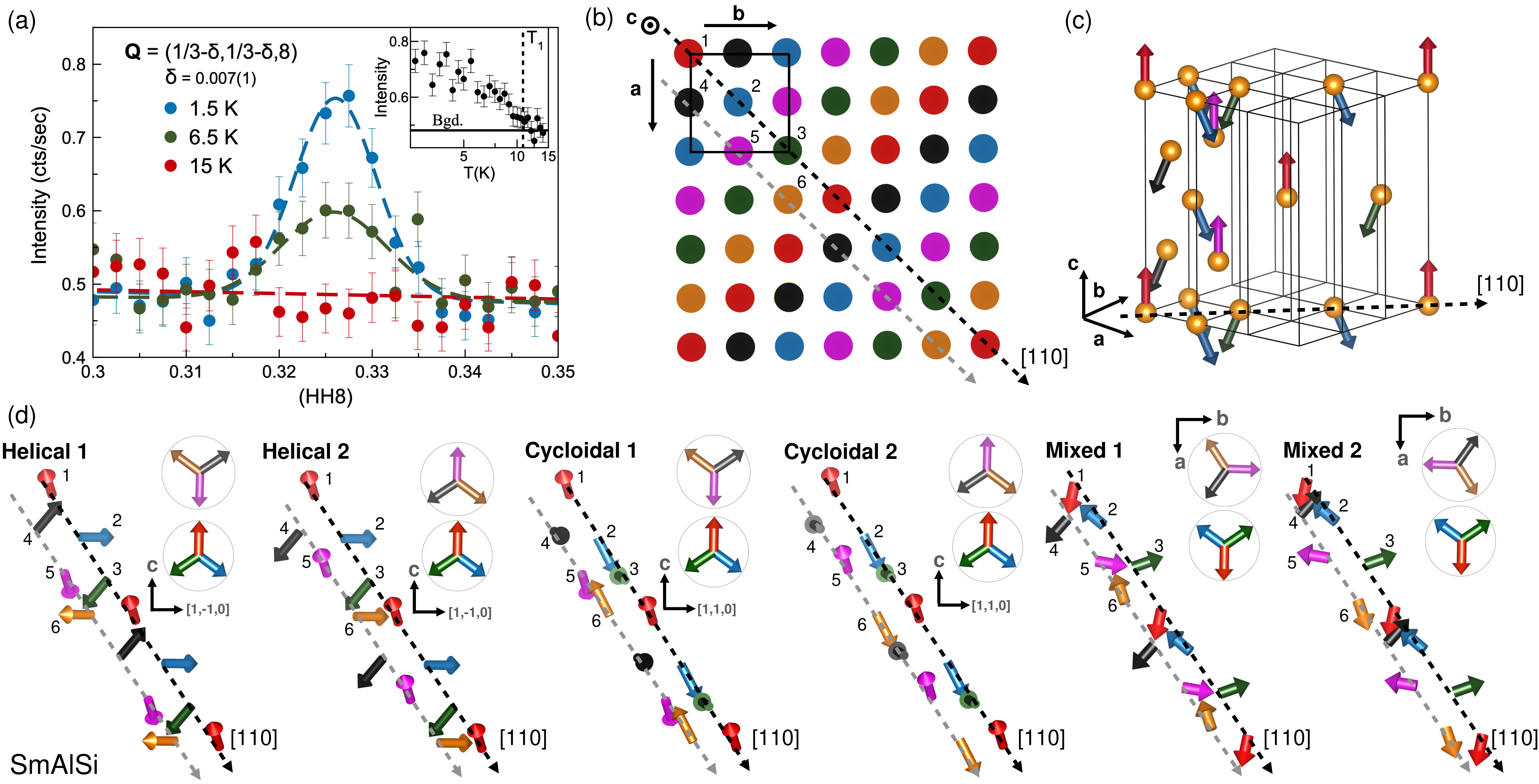}
	\caption{\label{fig:NEUTRON}
	(a) Neutron diffraction scan along the $(HH8)$ direction in SmAlSi, which reveals a temperature dependent Bragg peak at $\textbf{Q}=\left(\frac{1}{3}-\delta,\frac{1}{3}-\delta,8 \right)$. 
	The top right inset is the temperature dependence of the Bragg peak intensity. 
	Error bars correspond to one standard deviation.
	(b) View of the magnetic unit cell for a $\mathbf{k}=\left( \frac{1}{3},\frac{1}{3},0 \right)$ ordering vector. 
	Each color represents a different spin orientation on a particular Sm site. 
	The black box represents the size of the structural unit cell. 
	(c) 3D rendition of one possible spin structure. 
	(d) Sketches of all six possible spiral orders in SmAlSi. 
	The different colors represent different spin orientations as in (b).
	}
\end{figure*}

\emph{Neutron Diffraction.} It is challenging to probe the magnetic ground-state of SmAlSi with neutrons because the naturally occurring isotope $^{149}$Sm has a large neutron absorption cross-section ($\sigma_{\text{abs}}=42,080$~b) and Sm$^{3+}$ has a small moment ($0.7~\mu_B$).
Using $^{154}$Sm ($\sigma_{\text{abs}}=8.4$~b) could help but this isotope is available only in oxide form, which is not suitable for growing intermetallic crystals.
In the absence of isotope-rich samples, we took advantage of the recent instrumentation upgrade of the HB-1A spectrometer where the beam flux has been increased significantly, and aligned a single crystal of SmAlSi within the (HHL) plane to probe the Bragg peaks along the high symmetry directions of a few Brillouin zone (BZ) centers.
The intensity of the nuclear Bragg peaks did not increase below $T_1$ or $T_2$ ruling out a $\mathbf{k}=\left(0,0,0\right)$ ferromagnetic (FM) order. 
Instead, we found antiferromagnetic (AFM) Bragg peaks at $\mathbf{Q}=\left( \frac{1}{3}-\delta,\frac{1}{3}-\delta,4 \right)$ and $\left( \frac{1}{3}-\delta,\frac{1}{3}-\delta,8 \right)$, which can be indexed by an incommensurate vector $\mathbf{k}=\left( \frac{1}{3}-\delta,\frac{1}{3}-\delta,0 \right)$ with $\delta=0.007(1)$ (Fig.~\ref{fig:NEUTRON}a).
The temperature dependence of an AFM peak intensity in the inset of Fig.~\ref{fig:NEUTRON}a shows the onset of order parameter at $T_1=10.6(2)$~K.

The magnetic ordering vector $\mathbf{k}=\left( \frac{1}{3}-\delta,\frac{1}{3}-\delta,0 \right)$ in SmAlSi is similar to the incommensurate ordering vector observed in NdAlSi~\cite{gaudet_weyl-mediated_2021}, and indicates an SDW propagating along the $[1,1,0]$ direction. 
However, a large Ising MCA in NdAlSi prevents spiral ordering and drives a ferrimagnetic up-down-down order with a finite FM component~\cite{gaudet_weyl-mediated_2021}, whereas a weak MCA in SmAlSi (Fig.~\ref{fig:CEF}d) makes the spiral ordering possible without an FM component.

By performing a magnetic representation analysis~\cite{SARAh}, we identified six possible spin structures that yield a $\mathbf{k}=\left( \frac{1}{3}-\delta,\frac{1}{3}-\delta,0 \right)$ vector in SmAlSi and have constant magnetization on each Sm site.
We plot the six possible spin structures in Figs.~\ref{fig:NEUTRON}b,c,d assuming $\delta=0$ (ideal 120$^{\circ}$ order) for a simple visualization. 
Their magnetic unit cell is $3\times3\times1$ times the structural unit cell and contains up to 6 different spins in an ABC-type structure propagating amongst both the $\left(0,0,0\right)$ and $\left(0,\frac{1}{2},\frac{1}{4}\right)$ Sm sites as shown in 2D and 3D, respectively, in Fig.~\ref{fig:NEUTRON}b and \ref{fig:NEUTRON}c.
All six structures in Fig.~\ref{fig:NEUTRON}d are spiral but with different chirality. 
First, there is a helical order where the spins are constrained in the plane perpendicular to the $[1,1,0]$ propagation vector.
Second, there is a cycloidal order where the spins are constrained in the plane parallel to the $[1,1,0]$ propagation vector. 
Finally, there is a mixed spiral order where the spins are constrained within the basal plane so they have a component that oscillates both perpendicular and parallel to the propagation vector. 
There are two copies of each type of spiral with in-phase and anti-phase ordering between the $(0,0,0)$ and $\left(0,\frac{1}{2},\frac{1}{4}\right)$ primitive Sm sites as shown in Fig.~\ref{fig:NEUTRON}d.
Of these six possible configurations, the first two (helical 1 and 2) are consistent with the theoretical prediction that the $\mathbf{D}$ vector in the Weyl-induced DM interaction is (nearly) parallel to the bond direction when Weyl electrons have (nearly) spherically symmetric dispersion~\cite{nikolic_dynamics_2021}.

\begin{figure*}
	\includegraphics[width=\textwidth]{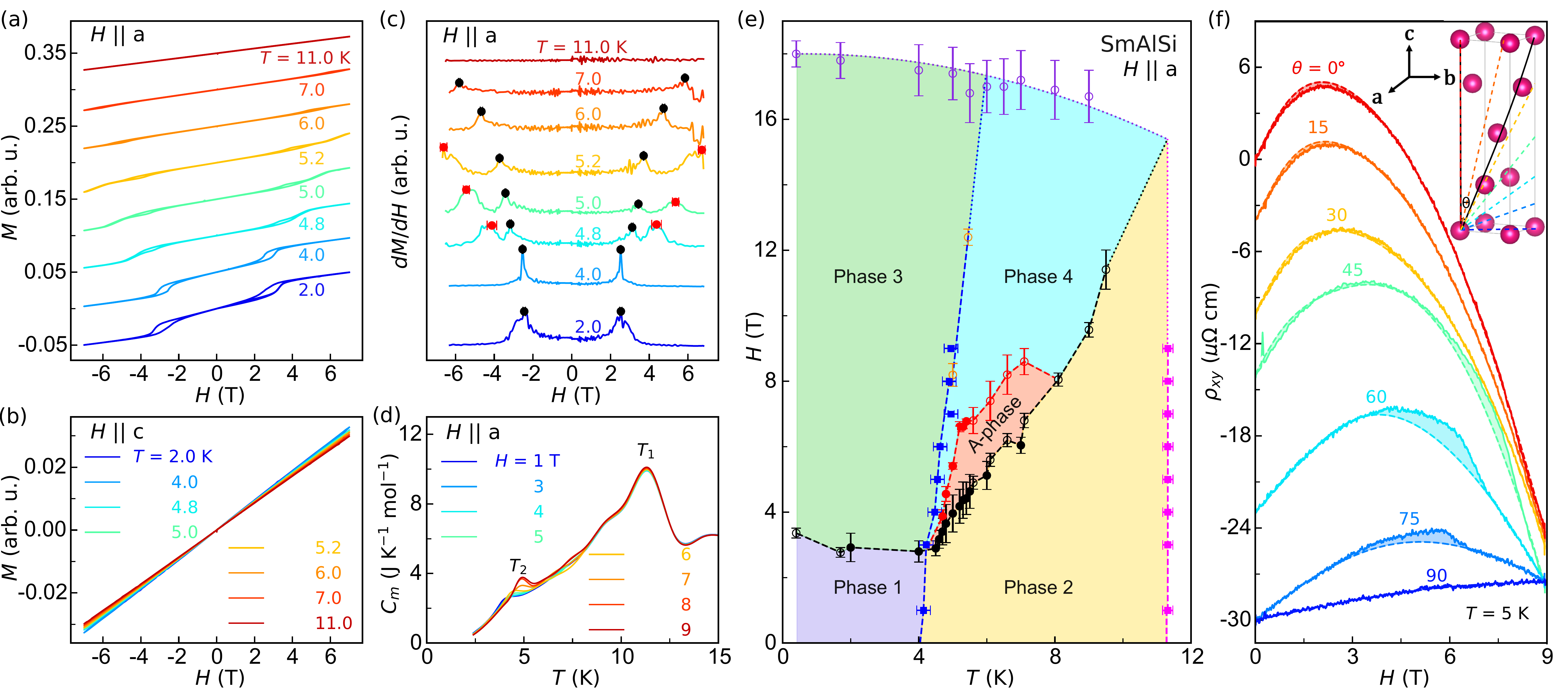}
	\caption{\label{fig:PD}
	(a) Magnetization curves with $H\|a$ are plotted at several temperatures. The 2~K curve has correct values in units of $\mu_B$ and the rest of the curves are shifted uniformly to improve visibility. 
	(b) $M(H)$ curves with $H\|c$ are strictly linear.
	(c) Derivatives of the $M(H\|a)$ curves reveal two sets of peaks (black and red circles) that define the A-phase boundaries.
	(d) Magnetic heat capacity as a function of temperature $C_m(T)$ at several fields with $H\|a$.
	(e) The phase diagram of SmAlSi. The filled magenta and blue squares correspond to $T_1$ and $T_2$, respectively. The filled black and red circles are from the panel (c). All empty symbols are from the high-field data in Fig.~S3~\cite{suppmatt}.
	(f) The Hall data $\rho_{xy}(H)$ are shown at 5~K with the field pointing at several angles as indicated in the inset.
	}
\end{figure*}

\emph{Magnetization.} In all spiral configurations of Fig.~\ref{fig:NEUTRON}d, the spins rotate 120$^\circ$ with respect to each other, such that no net magnetization is produced over an ABC cycle.
For SmAlSi, however, the spins rotate with 120.24(4)$^\circ$ with respect to each other due to the small incommensurability ($\delta$), so the magnetic unit cell is about 263~nm long with no net magnetization.
This is consistent with our $M(H)$ curves in Figs.~\ref{fig:PD}a,b that show the absence of hysteresis near zero-field for both $H\|a$ and $H\|c$.
At finite fields, however, we observe hysteretic metamagnetic transitions when $H\|a$ (Fig.~\ref{fig:PD}a) but not when $H\|c$ (Fig.~\ref{fig:PD}b).
The magnitude of $M$($H=7$~T) is only $0.045$~$\mu_B$ which is an order of magnitude less than the expected value for Sm$^{3+}$ (0.7~$\mu_B$).
The hysteretic metamagnetic transitions with small amplitudes for $H\|a$ and the strictly linear $M(H)$ curves when $H\|c$ are characteristics of materials with topological skyrmion excitations, e.g. GaV$_4$(S,Se)$_8$ and Gd$_2$PdSi$_3$~\cite{ruff_multiferroicity_nodate,fujima_thermodynamically_2017,bordacs_equilibrium_2017,kurumaji_skyrmion_2019}. 

We construct a magnetic phase diagram for SmAlSi based on two sets of peaks in $dM/dH$ as a function of field (black and red circles in Fig.~\ref{fig:PD}c) and the field dependence of $T_1$ and $T_2$ peaks in the $C_m$ data (Fig.~\ref{fig:PD}d).
There are five phases in the phase diagram (Fig.~\ref{fig:PD}e).
Phase 1 is the closest region to the zero-field ground-state investigated by neutron diffraction (Fig.~\ref{fig:NEUTRON}).
Its horizontal and vertical boundaries correspond, respectively, to the black peaks in $dM/dH$ and $T_2$ peaks in the $C_m$ data (Figs.~\ref{fig:PD}c,d).
Phase 2 is bound vertically by the $T_1$ and $T_2$ peaks in the $C_m$ data (Fig.~\ref{fig:PD}d).
It is bound horizontally by the black peaks in $dM/dH$ (filled black circles in Fig.~\ref{fig:PD}c,e) and hysteresis loops in the high-field magnetic torque data (empty black circles in Fig.~\ref{fig:PD}e).
The high-field data are presented as Supplemental Material~\cite{suppmatt} (Fig.~S3).
Phase 3 is horizontally bound by the black peaks in $dM/dH$ and purple circles representing the saturation of the high-field torque signal (Fig.~S3).
It is vertically bound by the $T_2$ peaks in the $C_m$ data (blue squares in Fig.~\ref{fig:PD}e) and steps in the high-field torque data (open orange circles)~\cite{suppmatt}.
The two sets of $dM/dH$ peaks (black and red symbols in Fig.~\ref{fig:PD}c,e) encompass an A-phase similar to the A-phase in materials such as GaV$_4$(S,Se)$_8$ and Gd$_2$PdSi$_3$ where skyrmions have been found~\cite{ruff_multiferroicity_nodate,fujima_thermodynamically_2017,bordacs_equilibrium_2017,kurumaji_skyrmion_2019}.
Phase 4 is separated from the A-phase by the red $dM/dH$ peaks in Fig.~\ref{fig:PD}c and hysteresis loops in the high-field torque data (open red circles in Fig.~\ref{fig:PD}e and Fig.~S3).

\emph{Hall Effect.} Confirming skyrmions in the A-phase of SmAlSi is highly challenging given that isotope-rich samples (with a small neutron absorption cross-section) are currently unavailable.
However, we detect a topological Hall signal $\rho_{xy}^T$ within the temperature and field range of the A-phase that indicates a multi-$q$ non-collinear spin texture as observed in skyrmion lattices~\cite{kurumaji_skyrmion_2019,leroux_skyrmion_2018}.
Figure~\ref{fig:PD}f shows the total Hall signal $\rho_{xy}=\rho_{xy}^O+\rho_{xy}^T$ measured with the field directed at different angles ($\theta$) with respect to the unit cell as indicated in the inset.
The underlying nonlinear field dependence of the ordinary Hall signal $\rho_{xy}^O(H)$ results from multiband conduction~\cite{lin_multiband_2016}, which is fitted to a three-band model (dashed lines in Fig.~\ref{fig:PD}f).
Details of the three-band model are discussed in the Supplemental Material~\cite{suppmatt} (Fig.~S4).
We highlight the topological Hall signal $\rho_{xy}^T = \rho_{xy} - \rho_{xy}^O$ with colors in Fig.~\ref{fig:PD}f.
Note that $\rho_{xy}^T$ is observed at $T=5$~K between 4 and 7~T, i.e. within the A-phase, and it is maximized when the field is nearly aligned with the body diagonal of the magnetic unit cell ($45^\circ \le \theta \le 75^\circ$).
It is lost at temperatures lower or higher than the A-phase boundaries.
The emergence of an A-phase is facilitated by a Weyl-induced three-spin ``chiral'' interaction ${\bf S}_1 \cdot({\bf S}_2 \times {\bf S}_3)$ whose coupling strength is proportional to the applied magnetic field \cite{nikolic_dynamics_2021}. 
Since the chiral interaction acts as a chemical potential for skyrmions \cite{nikolic_quantum_2020}, a finite field and/or a finite temperature (which destabilizes conventional magnetic order) are needed for the A-phase.

\emph{Electronic Structure.}
\begin{figure}
	\includegraphics[width=0.48\textwidth]{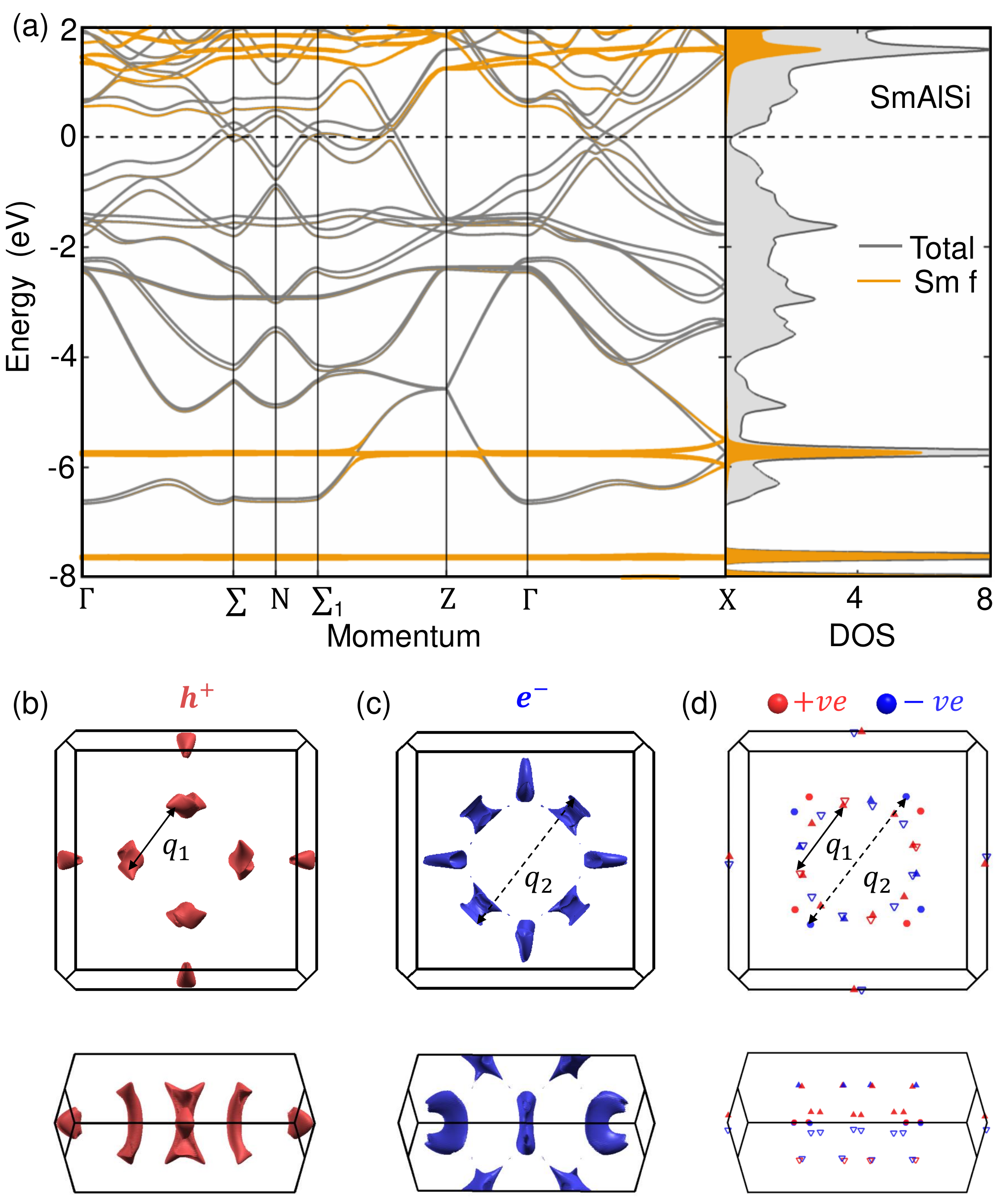}
	\caption{\label{fig:DFT}
	(a) Band structure and density of states in SmAlSi where $E_F$ is marked by a dashed line. The $f$-bands are highlighted in orange color. 
	(b) Top and side views of the hole pocket in the first BZ.
	(c) Same views of the electron pocket.
	(d) A map of the Weyl nodes where the circles, full triangles, and empty triangles represent Weyl nodes at $k_z=0$, $k_z>0$, and $k_z<0$, respectively.
	}
\end{figure}
The observation of a spiral magnetic ground-state with a possible skyrmion phase in SmAlSi indicates sizable DM and chiral interactions. 
Recent calculations have shown that Weyl electrons could mediate such interactions by coupling to the local $f$-moments through an RKKY mechanism~\cite{nikolic_universal_2021,nikolic_dynamics_2021,wang_rkky_2017,chang_rkky_2015}.
The band structure of SmAlSi in Fig.~\ref{fig:DFT}a has the necessary ingredients of the Weyl-mediated RKKY mechanism, namely (i) localized $f$-bands well below the Fermi level ($E_F$), (ii) a small density of states (DOS) near $E_F$ from the linearly dispersing bands with Weyl crossings, and (iii) intra-band $k$-space vectors (nesting vectors) that match the magnetic ordering vectors and link topologically non-trivial Fermi pockets.
To verify the last point, we visualize the Fermi surface of SmAlSi from DFT+SOC+$U$ calculations in the field-driven FM state.
It comprises one hole and one electron pocket (Figs.~\ref{fig:DFT}b,c) and 20 pairs of Weyl nodes (Fig.~\ref{fig:DFT}d). 
Details of the DFT calculations are presented in the Supplemental Material\cite{suppmatt}, where the Weyl nodes are categorized into 5 groups ($W_1$ to $W_5$) based on their symmetries and energies (Figs.~S5 and S6).
The closest and farthest Weyl nodes with respect to $E_F$ are $W_1$ and $W_3$ nodes at $E=0$ and $89$~meV, respectively (Fig.~S6).

As shown in Fig.~\ref{fig:DFT}b,c, two momentum space vectors $\mathbf{q_1} =\left( \frac{1}{3}-\delta,\frac{1}{3}-\delta,0 \right)$ and $\mathbf{q_2}=\left( \frac{2}{3}+\delta,\frac{2}{3}+\delta,0 \right)$ connect topologically non-trivial Fermi pockets. 
Both nesting vectors are consistent with the observed spiral ordering vector $\mathbf{k}=\left( \frac{1}{3}-\delta,\frac{1}{3}-\delta,0 \right)$ in Fig.~\ref{fig:NEUTRON}.
Specifically, $\mathbf{q_1}$ and $\mathbf{q_2}$ connect the $W_1$ and $W_5$ nodes at $E=0$ and $23$~meV, respectively.
The nesting effect in SmAlSi is enhanced by the elongation of both the hole and electron pockets along the $L$-direction (Figs.~\ref{fig:DFT}b,c).
We confirmed the accuracy of our DFT calculations by comparing the theoretical dimensions of the Fermi pockets to the frequency of the quantum oscillations in Fig.~S7~\cite{suppmatt,rourke_numerical_2012}.

Although similar nesting conditions can be found in the band structures of CeAlSi, PrAlSi, and NdAlSi (Fig.~S5)~\cite{yang_noncollinear_2021,yang_transition_2020,gaudet_weyl-mediated_2021}, the strong MCA blocks spiral ordering in those systems.
PrAlSi has an out-of-plane FM order while CeAlSi has an in-plane non-collinear FM order, and neither one shows evidence of DM interactions~\cite{yang_noncollinear_2021}.
NdAlSi has an incommensurate AFM order at intermediate temperatures whose ordering vectors match the $\mathbf{q_1}$ and $\mathbf{q_2}$ nesting vectors~\cite{gaudet_weyl-mediated_2021}.
The magnetic Bragg peaks associated with the $\mathbf{q_1}$ and $\mathbf{q_2}$ vectors were respectively refined to a weak in-plane helical spin component and a strong out-of-plane Ising component with a net FM moment.
The Ising character of the NdAlSi incommensurate order caused by a strong MCA dominates the Weyl-mediated RKKY interaction, and upon cooling, a commensurate ferrimagnetic ground-state is ultimately reached.
For SmAlSi, however, our observations of an incommensurate magnetism persisting down to the lowest temperatures, a lack of ferromagnetism in zero field, and a quasi-isotropic magnetocrystalline interaction suggest that a perfect AFM spiral order could be stabilized by Weyl-mediated RKKY interactions.

\section{\label{sec:conclue}Conclusion}
We found a spiral magnetic order in the Weyl semimetal SmAlSi whose periodicity is linked to the distance in momentum space between Weyl Fermi pockets. 
Reminiscent of the A-phase observed in skyrmion lattices, we observed a topological Hall effect in SmAlSi that appears only at finite fields and temperatures close to the onset of the magnetic order.
Our observations are thus fully consistent with Weyl-mediated RKKY interactions and in-field chiral three-spin interactions that promote Skyrmion lattices. 
For completeness, we present a Monte Carlo (MC) simulation of the Weyl-mediated RKKY interactions as Supplemental Material~\cite{suppmatt} based on the structural parameters of SmAlSi and reasonable estimates of the coupling constants.
Our MC simulations (Figs.~S8, S9, S10) provide a proof-of-principle demonstration of how Weyl-mediate RKKY interactions lead to a spiral order in the ground-state (consistent with the observed wave vector) and a multi-$q$ magnetic order at finite temperatures (consistent with the observed A-phase).
In the absence of isotope-rich samples, our neutron diffraction data does not resolve all details of the magnetic structure, especially at temperature between $T_1$ and $T_2$.



\section*{ACKNOWLEDGMENTS}
This material is based upon work supported by the Air Force Office of Scientific Research under award number FA2386-21-1-4059.
S.M.H. is supported by the MOST-AFOSR Taiwan program on Topological and Nanostructured Materials, Grant No. 110-2124-M-110-002-MY3
A portion of this research used resources at the High Flux Isotope Reactor, a DOE Office of Science User Facility operated by Oak Ridge National Laboratory. 
Any mention of commercial products is intended solely for fully detailing experiments; it does not imply recommendation or endorsement by NIST.
A portion of this work was performed at the National High Magnetic Field Laboratory, which is supported by the National Science Foundation Cooperative Agreement no. DMR-1644779 and the state of Florida. P.N. acknowledges support from the Institute for Quantum Matter, an Energy Frontier Research Center funded by the U.S. Department of Energy, Office of Science, Basic Energy Sciences under Award No. DE-SC0019331.
The work at Northeastern University was supported by the US Department of Energy (DOE), Office of Science, Basic Energy Sciences Grant No. DE-SC0022216 and benefited from Northeastern University’s Advanced Scientific Computation Center and the Discovery Cluster and the National Energy Research Scientific Computing Center through DOE Grant No. DE-AC02-05CH11231.
The work at TIFR Mumbai was supported by the Department of Atomic Energy of the Government of India under Project No. 12-R$\&$D-TFR-5.10-0100.

\bibliography{Yao_3June2022}

\begin{thebibliography}{36}%
\makeatletter
\providecommand \@ifxundefined [1]{%
 \@ifx{#1\undefined}
}%
\providecommand \@ifnum [1]{%
 \ifnum #1\expandafter \@firstoftwo
 \else \expandafter \@secondoftwo
 \fi
}%
\providecommand \@ifx [1]{%
 \ifx #1\expandafter \@firstoftwo
 \else \expandafter \@secondoftwo
 \fi
}%
\providecommand \natexlab [1]{#1}%
\providecommand \enquote  [1]{``#1''}%
\providecommand \bibnamefont  [1]{#1}%
\providecommand \bibfnamefont [1]{#1}%
\providecommand \citenamefont [1]{#1}%
\providecommand \href@noop [0]{\@secondoftwo}%
\providecommand \href [0]{\begingroup \@sanitize@url \@href}%
\providecommand \@href[1]{\@@startlink{#1}\@@href}%
\providecommand \@@href[1]{\endgroup#1\@@endlink}%
\providecommand \@sanitize@url [0]{\catcode `\\12\catcode `\$12\catcode
  `\&12\catcode `\#12\catcode `\^12\catcode `\_12\catcode `\%12\relax}%
\providecommand \@@startlink[1]{}%
\providecommand \@@endlink[0]{}%
\providecommand \url  [0]{\begingroup\@sanitize@url \@url }%
\providecommand \@url [1]{\endgroup\@href {#1}{\urlprefix }}%
\providecommand \urlprefix  [0]{URL }%
\providecommand \Eprint [0]{\href }%
\providecommand \doibase [0]{http://dx.doi.org/}%
\providecommand \selectlanguage [0]{\@gobble}%
\providecommand \bibinfo  [0]{\@secondoftwo}%
\providecommand \bibfield  [0]{\@secondoftwo}%
\providecommand \translation [1]{[#1]}%
\providecommand \BibitemOpen [0]{}%
\providecommand \bibitemStop [0]{}%
\providecommand \bibitemNoStop [0]{.\EOS\space}%
\providecommand \EOS [0]{\spacefactor3000\relax}%
\providecommand \BibitemShut  [1]{\csname bibitem#1\endcsname}%
\let\auto@bib@innerbib\@empty
\bibitem [{\citenamefont {Tokura}\ \emph {et~al.}(2019)\citenamefont {Tokura},
  \citenamefont {Yasuda},\ and\ \citenamefont
  {Tsukazaki}}]{tokura_magnetic_2019}%
  \BibitemOpen
  \bibfield  {author} {\bibinfo {author} {\bibfnamefont {Y.}~\bibnamefont
  {Tokura}}, \bibinfo {author} {\bibfnamefont {K.}~\bibnamefont {Yasuda}}, \
  and\ \bibinfo {author} {\bibfnamefont {A.}~\bibnamefont {Tsukazaki}},\ }\href
  {\doibase 10.1038/s42254-018-0011-5} {\bibfield  {journal} {\bibinfo
  {journal} {Nature Reviews Physics}\ }\textbf {\bibinfo {volume} {1}},\
  \bibinfo {pages} {126} (\bibinfo {year} {2019})}\BibitemShut {NoStop}%
\bibitem [{\citenamefont {Wang}\ \emph
  {et~al.}(2021{\natexlab{a}})\citenamefont {Wang}, \citenamefont {Rogers},
  \citenamefont {Yao}, \citenamefont {Nichols}, \citenamefont {Atay},
  \citenamefont {Xu}, \citenamefont {Franklin}, \citenamefont {Sochnikov},
  \citenamefont {Ryan}, \citenamefont {Haskel},\ and\ \citenamefont
  {Tafti}}]{wang_colossal_2021}%
  \BibitemOpen
  \bibfield  {author} {\bibinfo {author} {\bibfnamefont {Z.-C.}\ \bibnamefont
  {Wang}}, \bibinfo {author} {\bibfnamefont {J.~D.}\ \bibnamefont {Rogers}},
  \bibinfo {author} {\bibfnamefont {X.}~\bibnamefont {Yao}}, \bibinfo {author}
  {\bibfnamefont {R.}~\bibnamefont {Nichols}}, \bibinfo {author} {\bibfnamefont
  {K.}~\bibnamefont {Atay}}, \bibinfo {author} {\bibfnamefont {B.}~\bibnamefont
  {Xu}}, \bibinfo {author} {\bibfnamefont {J.}~\bibnamefont {Franklin}},
  \bibinfo {author} {\bibfnamefont {I.}~\bibnamefont {Sochnikov}}, \bibinfo
  {author} {\bibfnamefont {P.~J.}\ \bibnamefont {Ryan}}, \bibinfo {author}
  {\bibfnamefont {D.}~\bibnamefont {Haskel}}, \ and\ \bibinfo {author}
  {\bibfnamefont {F.}~\bibnamefont {Tafti}},\ }\href {\doibase
  10.1002/adma.202005755} {\bibfield  {journal} {\bibinfo  {journal} {Advanced
  Materials}\ }\textbf {\bibinfo {volume} {33}},\ \bibinfo {pages} {2005755}
  (\bibinfo {year} {2021}{\natexlab{a}})}\BibitemShut {NoStop}%
\bibitem [{\citenamefont {Bestwick}\ \emph {et~al.}(2015)\citenamefont
  {Bestwick}, \citenamefont {Fox}, \citenamefont {Kou}, \citenamefont {Pan},
  \citenamefont {Wang},\ and\ \citenamefont
  {Goldhaber-Gordon}}]{bestwick_precise_2015}%
  \BibitemOpen
  \bibfield  {author} {\bibinfo {author} {\bibfnamefont {A.}~\bibnamefont
  {Bestwick}}, \bibinfo {author} {\bibfnamefont {E.}~\bibnamefont {Fox}},
  \bibinfo {author} {\bibfnamefont {X.}~\bibnamefont {Kou}}, \bibinfo {author}
  {\bibfnamefont {L.}~\bibnamefont {Pan}}, \bibinfo {author} {\bibfnamefont
  {K.~L.}\ \bibnamefont {Wang}}, \ and\ \bibinfo {author} {\bibfnamefont
  {D.}~\bibnamefont {Goldhaber-Gordon}},\ }\href {\doibase
  10.1103/PhysRevLett.114.187201} {\bibfield  {journal} {\bibinfo  {journal}
  {Physical Review Letters}\ }\textbf {\bibinfo {volume} {114}},\ \bibinfo
  {pages} {187201} (\bibinfo {year} {2015})}\BibitemShut {NoStop}%
\bibitem [{\citenamefont {Hu}\ \emph {et~al.}()\citenamefont {Hu},
  \citenamefont {Ding}, \citenamefont {Gordon}, \citenamefont {Ghosh},
  \citenamefont {Tien}, \citenamefont {Li}, \citenamefont {Linn}, \citenamefont
  {Lien}, \citenamefont {Huang}, \citenamefont {Mackey}, \citenamefont {Liu},
  \citenamefont {Reddy}, \citenamefont {Singh}, \citenamefont {Agarwal},
  \citenamefont {Bansil}, \citenamefont {Song}, \citenamefont {Li},
  \citenamefont {Xu}, \citenamefont {Lin}, \citenamefont {Cao}, \citenamefont
  {Chang}, \citenamefont {Dessau},\ and\ \citenamefont
  {Ni}}]{hu_realization_nodate}%
  \BibitemOpen
  \bibfield  {author} {\bibinfo {author} {\bibfnamefont {C.}~\bibnamefont
  {Hu}}, \bibinfo {author} {\bibfnamefont {L.}~\bibnamefont {Ding}}, \bibinfo
  {author} {\bibfnamefont {K.~N.}\ \bibnamefont {Gordon}}, \bibinfo {author}
  {\bibfnamefont {B.}~\bibnamefont {Ghosh}}, \bibinfo {author} {\bibfnamefont
  {H.-J.}\ \bibnamefont {Tien}}, \bibinfo {author} {\bibfnamefont
  {H.}~\bibnamefont {Li}}, \bibinfo {author} {\bibfnamefont {A.~G.}\
  \bibnamefont {Linn}}, \bibinfo {author} {\bibfnamefont {S.-W.}\ \bibnamefont
  {Lien}}, \bibinfo {author} {\bibfnamefont {C.-Y.}\ \bibnamefont {Huang}},
  \bibinfo {author} {\bibfnamefont {S.}~\bibnamefont {Mackey}}, \bibinfo
  {author} {\bibfnamefont {J.}~\bibnamefont {Liu}}, \bibinfo {author}
  {\bibfnamefont {P.~V.~S.}\ \bibnamefont {Reddy}}, \bibinfo {author}
  {\bibfnamefont {B.}~\bibnamefont {Singh}}, \bibinfo {author} {\bibfnamefont
  {A.}~\bibnamefont {Agarwal}}, \bibinfo {author} {\bibfnamefont
  {A.}~\bibnamefont {Bansil}}, \bibinfo {author} {\bibfnamefont
  {M.}~\bibnamefont {Song}}, \bibinfo {author} {\bibfnamefont {D.}~\bibnamefont
  {Li}}, \bibinfo {author} {\bibfnamefont {S.-Y.}\ \bibnamefont {Xu}}, \bibinfo
  {author} {\bibfnamefont {H.}~\bibnamefont {Lin}}, \bibinfo {author}
  {\bibfnamefont {H.}~\bibnamefont {Cao}}, \bibinfo {author} {\bibfnamefont
  {T.-R.}\ \bibnamefont {Chang}}, \bibinfo {author} {\bibfnamefont
  {D.}~\bibnamefont {Dessau}}, \ and\ \bibinfo {author} {\bibfnamefont
  {N.}~\bibnamefont {Ni}},\ }\href {\doibase 10.1126/sciadv.aba4275} {\bibfield
   {journal} {\bibinfo  {journal} {Science Advances}\ }\textbf {\bibinfo
  {volume} {6}},\ \bibinfo {pages} {eaba4275}}\BibitemShut {NoStop}%
\bibitem [{\citenamefont {Liu}\ \emph {et~al.}(2020)\citenamefont {Liu},
  \citenamefont {Wang}, \citenamefont {Li}, \citenamefont {Wu}, \citenamefont
  {Li}, \citenamefont {Li}, \citenamefont {He}, \citenamefont {Xu},
  \citenamefont {Zhang},\ and\ \citenamefont {Wang}}]{liu_robust_2020}%
  \BibitemOpen
  \bibfield  {author} {\bibinfo {author} {\bibfnamefont {C.}~\bibnamefont
  {Liu}}, \bibinfo {author} {\bibfnamefont {Y.}~\bibnamefont {Wang}}, \bibinfo
  {author} {\bibfnamefont {H.}~\bibnamefont {Li}}, \bibinfo {author}
  {\bibfnamefont {Y.}~\bibnamefont {Wu}}, \bibinfo {author} {\bibfnamefont
  {Y.}~\bibnamefont {Li}}, \bibinfo {author} {\bibfnamefont {J.}~\bibnamefont
  {Li}}, \bibinfo {author} {\bibfnamefont {K.}~\bibnamefont {He}}, \bibinfo
  {author} {\bibfnamefont {Y.}~\bibnamefont {Xu}}, \bibinfo {author}
  {\bibfnamefont {J.}~\bibnamefont {Zhang}}, \ and\ \bibinfo {author}
  {\bibfnamefont {Y.}~\bibnamefont {Wang}},\ }\href {\doibase
  10.1038/s41563-019-0573-3} {\bibfield  {journal} {\bibinfo  {journal} {Nature
  Materials}\ }\textbf {\bibinfo {volume} {19}},\ \bibinfo {pages} {522}
  (\bibinfo {year} {2020})}\BibitemShut {NoStop}%
\bibitem [{\citenamefont {Sun}\ \emph {et~al.}(2021)\citenamefont {Sun},
  \citenamefont {Lee}, \citenamefont {Yang}, \citenamefont {Torchinsky},
  \citenamefont {Tafti},\ and\ \citenamefont {Orenstein}}]{sun_mapping_2021}%
  \BibitemOpen
  \bibfield  {author} {\bibinfo {author} {\bibfnamefont {Y.}~\bibnamefont
  {Sun}}, \bibinfo {author} {\bibfnamefont {C.}~\bibnamefont {Lee}}, \bibinfo
  {author} {\bibfnamefont {H.-Y.}\ \bibnamefont {Yang}}, \bibinfo {author}
  {\bibfnamefont {D.~H.}\ \bibnamefont {Torchinsky}}, \bibinfo {author}
  {\bibfnamefont {F.}~\bibnamefont {Tafti}}, \ and\ \bibinfo {author}
  {\bibfnamefont {J.}~\bibnamefont {Orenstein}},\ }\href {\doibase
  10.1103/PhysRevB.104.235119} {\bibfield  {journal} {\bibinfo  {journal}
  {Physical Review B}\ }\textbf {\bibinfo {volume} {104}},\ \bibinfo {pages}
  {235119} (\bibinfo {year} {2021})}\BibitemShut {NoStop}%
\bibitem [{\citenamefont {Xu}\ \emph {et~al.}(2021{\natexlab{a}})\citenamefont
  {Xu}, \citenamefont {Franklin}, \citenamefont {Jayacody}, \citenamefont
  {Yang}, \citenamefont {Tafti},\ and\ \citenamefont
  {Sochnikov}}]{xu_back_2021}%
  \BibitemOpen
  \bibfield  {author} {\bibinfo {author} {\bibfnamefont {B.}~\bibnamefont
  {Xu}}, \bibinfo {author} {\bibfnamefont {J.}~\bibnamefont {Franklin}},
  \bibinfo {author} {\bibfnamefont {A.}~\bibnamefont {Jayacody}}, \bibinfo
  {author} {\bibfnamefont {H.-Y.}\ \bibnamefont {Yang}}, \bibinfo {author}
  {\bibfnamefont {F.}~\bibnamefont {Tafti}}, \ and\ \bibinfo {author}
  {\bibfnamefont {I.}~\bibnamefont {Sochnikov}},\ }\href {\doibase
  10.1002/qute.202170033} {\bibfield  {journal} {\bibinfo  {journal} {Advanced
  Quantum Technologies}\ }\textbf {\bibinfo {volume} {4}},\ \bibinfo {pages}
  {2170033} (\bibinfo {year} {2021}{\natexlab{a}})}\BibitemShut {NoStop}%
\bibitem [{\citenamefont
  {Nikolić}(2021{\natexlab{a}})}]{nikolic_universal_2021}%
  \BibitemOpen
  \bibfield  {author} {\bibinfo {author} {\bibfnamefont {P.}~\bibnamefont
  {Nikolić}},\ }\href {\doibase 10.1103/PhysRevB.104.024414} {\bibfield
  {journal} {\bibinfo  {journal} {Physical Review B}\ }\textbf {\bibinfo
  {volume} {104}},\ \bibinfo {pages} {024414} (\bibinfo {year}
  {2021}{\natexlab{a}})}\BibitemShut {NoStop}%
\bibitem [{\citenamefont
  {Nikolić}(2021{\natexlab{b}})}]{nikolic_dynamics_2021}%
  \BibitemOpen
  \bibfield  {author} {\bibinfo {author} {\bibfnamefont {P.}~\bibnamefont
  {Nikolić}},\ }\href {\doibase 10.1103/PhysRevB.103.155151} {\bibfield
  {journal} {\bibinfo  {journal} {Physical Review B}\ }\textbf {\bibinfo
  {volume} {103}},\ \bibinfo {pages} {155151} (\bibinfo {year}
  {2021}{\natexlab{b}})}\BibitemShut {NoStop}%
\bibitem [{\citenamefont {Wang}\ \emph {et~al.}(2017)\citenamefont {Wang},
  \citenamefont {Chang},\ and\ \citenamefont {Zhou}}]{wang_rkky_2017}%
  \BibitemOpen
  \bibfield  {author} {\bibinfo {author} {\bibfnamefont {S.-X.}\ \bibnamefont
  {Wang}}, \bibinfo {author} {\bibfnamefont {H.-R.}\ \bibnamefont {Chang}}, \
  and\ \bibinfo {author} {\bibfnamefont {J.}~\bibnamefont {Zhou}},\ }\href
  {\doibase 10.1103/PhysRevB.96.115204} {\bibfield  {journal} {\bibinfo
  {journal} {Physical Review B}\ }\textbf {\bibinfo {volume} {96}},\ \bibinfo
  {pages} {115204} (\bibinfo {year} {2017})}\BibitemShut {NoStop}%
\bibitem [{\citenamefont {Chang}\ \emph {et~al.}(2015)\citenamefont {Chang},
  \citenamefont {Zhou}, \citenamefont {Wang}, \citenamefont {Shan},\ and\
  \citenamefont {Xiao}}]{chang_rkky_2015}%
  \BibitemOpen
  \bibfield  {author} {\bibinfo {author} {\bibfnamefont {H.-R.}\ \bibnamefont
  {Chang}}, \bibinfo {author} {\bibfnamefont {J.}~\bibnamefont {Zhou}},
  \bibinfo {author} {\bibfnamefont {S.-X.}\ \bibnamefont {Wang}}, \bibinfo
  {author} {\bibfnamefont {W.-Y.}\ \bibnamefont {Shan}}, \ and\ \bibinfo
  {author} {\bibfnamefont {D.}~\bibnamefont {Xiao}},\ }\href {\doibase
  10.1103/PhysRevB.92.241103} {\bibfield  {journal} {\bibinfo  {journal}
  {Physical Review B}\ }\textbf {\bibinfo {volume} {92}},\ \bibinfo {pages}
  {241103} (\bibinfo {year} {2015})}\BibitemShut {NoStop}%
\bibitem [{\citenamefont {Wang}\ \emph
  {et~al.}(2021{\natexlab{b}})\citenamefont {Wang}, \citenamefont {Su},
  \citenamefont {Lin},\ and\ \citenamefont {Batista}}]{wang_meron_2021}%
  \BibitemOpen
  \bibfield  {author} {\bibinfo {author} {\bibfnamefont {Z.}~\bibnamefont
  {Wang}}, \bibinfo {author} {\bibfnamefont {Y.}~\bibnamefont {Su}}, \bibinfo
  {author} {\bibfnamefont {S.-Z.}\ \bibnamefont {Lin}}, \ and\ \bibinfo
  {author} {\bibfnamefont {C.~D.}\ \bibnamefont {Batista}},\ }\href {\doibase
  10.1103/PhysRevB.103.104408} {\bibfield  {journal} {\bibinfo  {journal}
  {Physical Review B}\ }\textbf {\bibinfo {volume} {103}},\ \bibinfo {pages}
  {104408} (\bibinfo {year} {2021}{\natexlab{b}})}\BibitemShut {NoStop}%
\bibitem [{\citenamefont {Repicky}\ \emph {et~al.}(2021)\citenamefont
  {Repicky}, \citenamefont {Wu}, \citenamefont {Liu}, \citenamefont {Corbett},
  \citenamefont {Zhu}, \citenamefont {Cheng}, \citenamefont {Ahmed},
  \citenamefont {Takeuchi}, \citenamefont {Guerrero-Sanchez}, \citenamefont
  {Randeria}, \citenamefont {Kawakami},\ and\ \citenamefont
  {Gupta}}]{repicky_atomic-scale_2021}%
  \BibitemOpen
  \bibfield  {author} {\bibinfo {author} {\bibfnamefont {J.}~\bibnamefont
  {Repicky}}, \bibinfo {author} {\bibfnamefont {P.-K.}\ \bibnamefont {Wu}},
  \bibinfo {author} {\bibfnamefont {T.}~\bibnamefont {Liu}}, \bibinfo {author}
  {\bibfnamefont {J.~P.}\ \bibnamefont {Corbett}}, \bibinfo {author}
  {\bibfnamefont {T.}~\bibnamefont {Zhu}}, \bibinfo {author} {\bibfnamefont
  {S.}~\bibnamefont {Cheng}}, \bibinfo {author} {\bibfnamefont {A.~S.}\
  \bibnamefont {Ahmed}}, \bibinfo {author} {\bibfnamefont {N.}~\bibnamefont
  {Takeuchi}}, \bibinfo {author} {\bibfnamefont {J.}~\bibnamefont
  {Guerrero-Sanchez}}, \bibinfo {author} {\bibfnamefont {M.}~\bibnamefont
  {Randeria}}, \bibinfo {author} {\bibfnamefont {R.~K.}\ \bibnamefont
  {Kawakami}}, \ and\ \bibinfo {author} {\bibfnamefont {J.~A.}\ \bibnamefont
  {Gupta}},\ }\href {\doibase 10.1126/science.abd9225} {\bibfield  {journal}
  {\bibinfo  {journal} {Science}\ }\textbf {\bibinfo {volume} {374}},\ \bibinfo
  {pages} {1484} (\bibinfo {year} {2021})}\BibitemShut {NoStop}%
\bibitem [{\citenamefont {Luo}\ and\ \citenamefont
  {You}(2021)}]{luo_skyrmion_2021}%
  \BibitemOpen
  \bibfield  {author} {\bibinfo {author} {\bibfnamefont {S.}~\bibnamefont
  {Luo}}\ and\ \bibinfo {author} {\bibfnamefont {L.}~\bibnamefont {You}},\
  }\href {\doibase 10.1063/5.0042917} {\bibfield  {journal} {\bibinfo
  {journal} {APL Materials}\ }\textbf {\bibinfo {volume} {9}},\ \bibinfo
  {pages} {050901} (\bibinfo {year} {2021})}\BibitemShut {NoStop}%
\bibitem [{\citenamefont {Gaudet}\ \emph {et~al.}(2021)\citenamefont {Gaudet},
  \citenamefont {Yang}, \citenamefont {Baidya}, \citenamefont {Lu},
  \citenamefont {Xu}, \citenamefont {Zhao}, \citenamefont {Rodriguez-Rivera},
  \citenamefont {Hoffmann}, \citenamefont {Graf}, \citenamefont {Torchinsky},
  \citenamefont {Nikolić}, \citenamefont {Vanderbilt}, \citenamefont {Tafti},\
  and\ \citenamefont {Broholm}}]{gaudet_weyl-mediated_2021}%
  \BibitemOpen
  \bibfield  {author} {\bibinfo {author} {\bibfnamefont {J.}~\bibnamefont
  {Gaudet}}, \bibinfo {author} {\bibfnamefont {H.-Y.}\ \bibnamefont {Yang}},
  \bibinfo {author} {\bibfnamefont {S.}~\bibnamefont {Baidya}}, \bibinfo
  {author} {\bibfnamefont {B.}~\bibnamefont {Lu}}, \bibinfo {author}
  {\bibfnamefont {G.}~\bibnamefont {Xu}}, \bibinfo {author} {\bibfnamefont
  {Y.}~\bibnamefont {Zhao}}, \bibinfo {author} {\bibfnamefont {J.~A.}\
  \bibnamefont {Rodriguez-Rivera}}, \bibinfo {author} {\bibfnamefont {C.~M.}\
  \bibnamefont {Hoffmann}}, \bibinfo {author} {\bibfnamefont {D.~E.}\
  \bibnamefont {Graf}}, \bibinfo {author} {\bibfnamefont {D.~H.}\ \bibnamefont
  {Torchinsky}}, \bibinfo {author} {\bibfnamefont {P.}~\bibnamefont
  {Nikolić}}, \bibinfo {author} {\bibfnamefont {D.}~\bibnamefont
  {Vanderbilt}}, \bibinfo {author} {\bibfnamefont {F.}~\bibnamefont {Tafti}}, \
  and\ \bibinfo {author} {\bibfnamefont {C.~L.}\ \bibnamefont {Broholm}},\
  }\href {\doibase 10.1038/s41563-021-01062-8} {\bibfield  {journal} {\bibinfo
  {journal} {Nature Materials}\ }\textbf {\bibinfo {volume} {20}},\ \bibinfo
  {pages} {1650} (\bibinfo {year} {2021})}\BibitemShut {NoStop}%
\bibitem [{\citenamefont {Neubauer}\ \emph {et~al.}(2009)\citenamefont
  {Neubauer}, \citenamefont {Pfleiderer}, \citenamefont {Binz}, \citenamefont
  {Rosch}, \citenamefont {Ritz}, \citenamefont {Niklowitz},\ and\ \citenamefont
  {Böni}}]{neubauer_topological_2009}%
  \BibitemOpen
  \bibfield  {author} {\bibinfo {author} {\bibfnamefont {A.}~\bibnamefont
  {Neubauer}}, \bibinfo {author} {\bibfnamefont {C.}~\bibnamefont
  {Pfleiderer}}, \bibinfo {author} {\bibfnamefont {B.}~\bibnamefont {Binz}},
  \bibinfo {author} {\bibfnamefont {A.}~\bibnamefont {Rosch}}, \bibinfo
  {author} {\bibfnamefont {R.}~\bibnamefont {Ritz}}, \bibinfo {author}
  {\bibfnamefont {P.~G.}\ \bibnamefont {Niklowitz}}, \ and\ \bibinfo {author}
  {\bibfnamefont {P.}~\bibnamefont {Böni}},\ }\href {\doibase
  10.1103/PhysRevLett.102.186602} {\bibfield  {journal} {\bibinfo  {journal}
  {Physical Review Letters}\ }\textbf {\bibinfo {volume} {102}},\ \bibinfo
  {pages} {186602} (\bibinfo {year} {2009})}\BibitemShut {NoStop}%
\bibitem [{\citenamefont {Hirschberger}\ \emph {et~al.}(2019)\citenamefont
  {Hirschberger}, \citenamefont {Nakajima}, \citenamefont {Gao}, \citenamefont
  {Peng}, \citenamefont {Kikkawa}, \citenamefont {Kurumaji}, \citenamefont
  {Kriener}, \citenamefont {Yamasaki}, \citenamefont {Sagayama}, \citenamefont
  {Nakao}, \citenamefont {Ohishi}, \citenamefont {Kakurai}, \citenamefont
  {Taguchi}, \citenamefont {Yu}, \citenamefont {Arima},\ and\ \citenamefont
  {Tokura}}]{hirschberger_skyrmion_2019}%
  \BibitemOpen
  \bibfield  {author} {\bibinfo {author} {\bibfnamefont {M.}~\bibnamefont
  {Hirschberger}}, \bibinfo {author} {\bibfnamefont {T.}~\bibnamefont
  {Nakajima}}, \bibinfo {author} {\bibfnamefont {S.}~\bibnamefont {Gao}},
  \bibinfo {author} {\bibfnamefont {L.}~\bibnamefont {Peng}}, \bibinfo {author}
  {\bibfnamefont {A.}~\bibnamefont {Kikkawa}}, \bibinfo {author} {\bibfnamefont
  {T.}~\bibnamefont {Kurumaji}}, \bibinfo {author} {\bibfnamefont
  {M.}~\bibnamefont {Kriener}}, \bibinfo {author} {\bibfnamefont
  {Y.}~\bibnamefont {Yamasaki}}, \bibinfo {author} {\bibfnamefont
  {H.}~\bibnamefont {Sagayama}}, \bibinfo {author} {\bibfnamefont
  {H.}~\bibnamefont {Nakao}}, \bibinfo {author} {\bibfnamefont
  {K.}~\bibnamefont {Ohishi}}, \bibinfo {author} {\bibfnamefont
  {K.}~\bibnamefont {Kakurai}}, \bibinfo {author} {\bibfnamefont
  {Y.}~\bibnamefont {Taguchi}}, \bibinfo {author} {\bibfnamefont
  {X.}~\bibnamefont {Yu}}, \bibinfo {author} {\bibfnamefont {T.-h.}\
  \bibnamefont {Arima}}, \ and\ \bibinfo {author} {\bibfnamefont
  {Y.}~\bibnamefont {Tokura}},\ }\href {\doibase 10.1038/s41467-019-13675-4}
  {\bibfield  {journal} {\bibinfo  {journal} {Nature Communications}\ }\textbf
  {\bibinfo {volume} {10}},\ \bibinfo {pages} {5831} (\bibinfo {year}
  {2019})}\BibitemShut {NoStop}%
\bibitem [{\citenamefont {Xu}\ \emph {et~al.}(2021{\natexlab{b}})\citenamefont
  {Xu}, \citenamefont {Niu}, \citenamefont {Bai}, \citenamefont {Zhu},
  \citenamefont {Yuan}, \citenamefont {He}, \citenamefont {Yang}, \citenamefont
  {Xia}, \citenamefont {Zhao},\ and\ \citenamefont
  {Tian}}]{xu_shubnikov-haas_2021}%
  \BibitemOpen
  \bibfield  {author} {\bibinfo {author} {\bibfnamefont {L.}~\bibnamefont
  {Xu}}, \bibinfo {author} {\bibfnamefont {H.}~\bibnamefont {Niu}}, \bibinfo
  {author} {\bibfnamefont {Y.}~\bibnamefont {Bai}}, \bibinfo {author}
  {\bibfnamefont {H.}~\bibnamefont {Zhu}}, \bibinfo {author} {\bibfnamefont
  {S.}~\bibnamefont {Yuan}}, \bibinfo {author} {\bibfnamefont {X.}~\bibnamefont
  {He}}, \bibinfo {author} {\bibfnamefont {Y.}~\bibnamefont {Yang}}, \bibinfo
  {author} {\bibfnamefont {Z.}~\bibnamefont {Xia}}, \bibinfo {author}
  {\bibfnamefont {L.}~\bibnamefont {Zhao}}, \ and\ \bibinfo {author}
  {\bibfnamefont {Z.}~\bibnamefont {Tian}},\ }\href {\doibase
  10.48550/arXiv.2107.11957} {\  (\bibinfo {year} {2021}{\natexlab{b}}),\
  10.48550/arXiv.2107.11957}\BibitemShut {NoStop}%
\bibitem [{\citenamefont
  {Rodr{\'\i}guez-Carvajal}(1993)}]{rodriguez-carvajal_recent_1993}%
  \BibitemOpen
  \bibfield  {author} {\bibinfo {author} {\bibfnamefont {J.}~\bibnamefont
  {Rodr{\'\i}guez-Carvajal}},\ }\href {\doibase 10.1016/0921-4526(93)90108-I}
  {\bibfield  {journal} {\bibinfo  {journal} {Physica B: Condensed Matter}\
  }\textbf {\bibinfo {volume} {192}},\ \bibinfo {pages} {55} (\bibinfo {year}
  {1993})}\BibitemShut {NoStop}%
\bibitem [{\citenamefont {Wills}(2000)}]{SARAh}%
  \BibitemOpen
  \bibfield  {author} {\bibinfo {author} {\bibfnamefont {A.}~\bibnamefont
  {Wills}},\ }\href {\doibase https://doi.org/10.1016/S0921-4526(99)01722-6}
  {\bibfield  {journal} {\bibinfo  {journal} {Physica B: Condensed Matter}\
  }\textbf {\bibinfo {volume} {276-278}},\ \bibinfo {pages} {680} (\bibinfo
  {year} {2000})}\BibitemShut {NoStop}%
\bibitem [{\citenamefont {Blaha}\ \emph {et~al.}(2020)\citenamefont {Blaha},
  \citenamefont {Schwarz}, \citenamefont {Tran}, \citenamefont {Laskowski},
  \citenamefont {Madsen},\ and\ \citenamefont {Marks}}]{blaha_wien2k_2020}%
  \BibitemOpen
  \bibfield  {author} {\bibinfo {author} {\bibfnamefont {P.}~\bibnamefont
  {Blaha}}, \bibinfo {author} {\bibfnamefont {K.}~\bibnamefont {Schwarz}},
  \bibinfo {author} {\bibfnamefont {F.}~\bibnamefont {Tran}}, \bibinfo {author}
  {\bibfnamefont {R.}~\bibnamefont {Laskowski}}, \bibinfo {author}
  {\bibfnamefont {G.~K.~H.}\ \bibnamefont {Madsen}}, \ and\ \bibinfo {author}
  {\bibfnamefont {L.~D.}\ \bibnamefont {Marks}},\ }\href {\doibase
  10.1063/1.5143061} {\bibfield  {journal} {\bibinfo  {journal} {The Journal of
  Chemical Physics}\ }\textbf {\bibinfo {volume} {152}},\ \bibinfo {pages}
  {074101} (\bibinfo {year} {2020})}\BibitemShut {NoStop}%
\bibitem [{\citenamefont {Perdew}\ \emph {et~al.}(1996)\citenamefont {Perdew},
  \citenamefont {Burke},\ and\ \citenamefont
  {Ernzerhof}}]{perdew_generalized_1996}%
  \BibitemOpen
  \bibfield  {author} {\bibinfo {author} {\bibfnamefont {J.~P.}\ \bibnamefont
  {Perdew}}, \bibinfo {author} {\bibfnamefont {K.}~\bibnamefont {Burke}}, \
  and\ \bibinfo {author} {\bibfnamefont {M.}~\bibnamefont {Ernzerhof}},\ }\href
  {\doibase 10.1103/PhysRevLett.77.3865} {\bibfield  {journal} {\bibinfo
  {journal} {Physical Review Letters}\ }\textbf {\bibinfo {volume} {77}},\
  \bibinfo {pages} {3865} (\bibinfo {year} {1996})}\BibitemShut {NoStop}%
\bibitem [{\citenamefont {Anisimov}\ \emph {et~al.}(1991)\citenamefont
  {Anisimov}, \citenamefont {Zaanen},\ and\ \citenamefont
  {Andersen}}]{anisimov_band_1991}%
  \BibitemOpen
  \bibfield  {author} {\bibinfo {author} {\bibfnamefont {V.~I.}\ \bibnamefont
  {Anisimov}}, \bibinfo {author} {\bibfnamefont {J.}~\bibnamefont {Zaanen}}, \
  and\ \bibinfo {author} {\bibfnamefont {O.~K.}\ \bibnamefont {Andersen}},\
  }\href {\doibase 10.1103/PhysRevB.44.943} {\bibfield  {journal} {\bibinfo
  {journal} {Physical Review B}\ }\textbf {\bibinfo {volume} {44}},\ \bibinfo
  {pages} {943} (\bibinfo {year} {1991})}\BibitemShut {NoStop}%
\bibitem [{\citenamefont {Lyu}\ \emph {et~al.}(2020)\citenamefont {Lyu},
  \citenamefont {Xiang}, \citenamefont {Mi}, \citenamefont {Zhao},
  \citenamefont {Wang}, \citenamefont {Liu}, \citenamefont {Chen},
  \citenamefont {Ren}, \citenamefont {Li},\ and\ \citenamefont
  {Sun}}]{lyu_nonsaturating_2020}%
  \BibitemOpen
  \bibfield  {author} {\bibinfo {author} {\bibfnamefont {M.}~\bibnamefont
  {Lyu}}, \bibinfo {author} {\bibfnamefont {J.}~\bibnamefont {Xiang}}, \bibinfo
  {author} {\bibfnamefont {Z.}~\bibnamefont {Mi}}, \bibinfo {author}
  {\bibfnamefont {H.}~\bibnamefont {Zhao}}, \bibinfo {author} {\bibfnamefont
  {Z.}~\bibnamefont {Wang}}, \bibinfo {author} {\bibfnamefont {E.}~\bibnamefont
  {Liu}}, \bibinfo {author} {\bibfnamefont {G.}~\bibnamefont {Chen}}, \bibinfo
  {author} {\bibfnamefont {Z.}~\bibnamefont {Ren}}, \bibinfo {author}
  {\bibfnamefont {G.}~\bibnamefont {Li}}, \ and\ \bibinfo {author}
  {\bibfnamefont {P.}~\bibnamefont {Sun}},\ }\href {\doibase
  10.1103/PhysRevB.102.085143} {\bibfield  {journal} {\bibinfo  {journal}
  {Physical Review B}\ }\textbf {\bibinfo {volume} {102}},\ \bibinfo {pages}
  {085143} (\bibinfo {year} {2020})}\BibitemShut {NoStop}%
\bibitem [{sup()}]{suppmatt}%
  \BibitemOpen
  \href {https://journals.aps.org} {}\bibinfo {note} {See the Supplemental
  Material for details}\BibitemShut {NoStop}%
\bibitem [{\citenamefont {Chang}\ \emph {et~al.}(2018)\citenamefont {Chang},
  \citenamefont {Singh}, \citenamefont {Xu}, \citenamefont {Bian},
  \citenamefont {Huang}, \citenamefont {Hsu}, \citenamefont {Belopolski},
  \citenamefont {Alidoust}, \citenamefont {Sanchez}, \citenamefont {Zheng},
  \citenamefont {Lu}, \citenamefont {Zhang}, \citenamefont {Bian},
  \citenamefont {Chang}, \citenamefont {Jeng}, \citenamefont {Bansil},
  \citenamefont {Hsu}, \citenamefont {Jia}, \citenamefont {Neupert},
  \citenamefont {Lin},\ and\ \citenamefont {Hasan}}]{chang_magnetic_2018}%
  \BibitemOpen
  \bibfield  {author} {\bibinfo {author} {\bibfnamefont {G.}~\bibnamefont
  {Chang}}, \bibinfo {author} {\bibfnamefont {B.}~\bibnamefont {Singh}},
  \bibinfo {author} {\bibfnamefont {S.-Y.}\ \bibnamefont {Xu}}, \bibinfo
  {author} {\bibfnamefont {G.}~\bibnamefont {Bian}}, \bibinfo {author}
  {\bibfnamefont {S.-M.}\ \bibnamefont {Huang}}, \bibinfo {author}
  {\bibfnamefont {C.-H.}\ \bibnamefont {Hsu}}, \bibinfo {author} {\bibfnamefont
  {I.}~\bibnamefont {Belopolski}}, \bibinfo {author} {\bibfnamefont
  {N.}~\bibnamefont {Alidoust}}, \bibinfo {author} {\bibfnamefont {D.~S.}\
  \bibnamefont {Sanchez}}, \bibinfo {author} {\bibfnamefont {H.}~\bibnamefont
  {Zheng}}, \bibinfo {author} {\bibfnamefont {H.}~\bibnamefont {Lu}}, \bibinfo
  {author} {\bibfnamefont {X.}~\bibnamefont {Zhang}}, \bibinfo {author}
  {\bibfnamefont {Y.}~\bibnamefont {Bian}}, \bibinfo {author} {\bibfnamefont
  {T.-R.}\ \bibnamefont {Chang}}, \bibinfo {author} {\bibfnamefont {H.-T.}\
  \bibnamefont {Jeng}}, \bibinfo {author} {\bibfnamefont {A.}~\bibnamefont
  {Bansil}}, \bibinfo {author} {\bibfnamefont {H.}~\bibnamefont {Hsu}},
  \bibinfo {author} {\bibfnamefont {S.}~\bibnamefont {Jia}}, \bibinfo {author}
  {\bibfnamefont {T.}~\bibnamefont {Neupert}}, \bibinfo {author} {\bibfnamefont
  {H.}~\bibnamefont {Lin}}, \ and\ \bibinfo {author} {\bibfnamefont {M.~Z.}\
  \bibnamefont {Hasan}},\ }\href {\doibase 10.1103/PhysRevB.97.041104}
  {\bibfield  {journal} {\bibinfo  {journal} {Physical Review B}\ }\textbf
  {\bibinfo {volume} {97}},\ \bibinfo {pages} {041104} (\bibinfo {year}
  {2018})}\BibitemShut {NoStop}%
\bibitem [{\citenamefont {Yang}\ \emph {et~al.}(2021)\citenamefont {Yang},
  \citenamefont {Singh}, \citenamefont {Gaudet}, \citenamefont {Lu},
  \citenamefont {Huang}, \citenamefont {Chiu}, \citenamefont {Huang},
  \citenamefont {Wang}, \citenamefont {Bahrami}, \citenamefont {Xu},
  \citenamefont {Franklin}, \citenamefont {Sochnikov}, \citenamefont {Graf},
  \citenamefont {Xu}, \citenamefont {Zhao}, \citenamefont {Hoffman},
  \citenamefont {Lin}, \citenamefont {Torchinsky}, \citenamefont {Broholm},
  \citenamefont {Bansil},\ and\ \citenamefont
  {Tafti}}]{yang_noncollinear_2021}%
  \BibitemOpen
  \bibfield  {author} {\bibinfo {author} {\bibfnamefont {H.-Y.}\ \bibnamefont
  {Yang}}, \bibinfo {author} {\bibfnamefont {B.}~\bibnamefont {Singh}},
  \bibinfo {author} {\bibfnamefont {J.}~\bibnamefont {Gaudet}}, \bibinfo
  {author} {\bibfnamefont {B.}~\bibnamefont {Lu}}, \bibinfo {author}
  {\bibfnamefont {C.-Y.}\ \bibnamefont {Huang}}, \bibinfo {author}
  {\bibfnamefont {W.-C.}\ \bibnamefont {Chiu}}, \bibinfo {author}
  {\bibfnamefont {S.-M.}\ \bibnamefont {Huang}}, \bibinfo {author}
  {\bibfnamefont {B.}~\bibnamefont {Wang}}, \bibinfo {author} {\bibfnamefont
  {F.}~\bibnamefont {Bahrami}}, \bibinfo {author} {\bibfnamefont
  {B.}~\bibnamefont {Xu}}, \bibinfo {author} {\bibfnamefont {J.}~\bibnamefont
  {Franklin}}, \bibinfo {author} {\bibfnamefont {I.}~\bibnamefont {Sochnikov}},
  \bibinfo {author} {\bibfnamefont {D.~E.}\ \bibnamefont {Graf}}, \bibinfo
  {author} {\bibfnamefont {G.}~\bibnamefont {Xu}}, \bibinfo {author}
  {\bibfnamefont {Y.}~\bibnamefont {Zhao}}, \bibinfo {author} {\bibfnamefont
  {C.~M.}\ \bibnamefont {Hoffman}}, \bibinfo {author} {\bibfnamefont
  {H.}~\bibnamefont {Lin}}, \bibinfo {author} {\bibfnamefont {D.~H.}\
  \bibnamefont {Torchinsky}}, \bibinfo {author} {\bibfnamefont {C.~L.}\
  \bibnamefont {Broholm}}, \bibinfo {author} {\bibfnamefont {A.}~\bibnamefont
  {Bansil}}, \ and\ \bibinfo {author} {\bibfnamefont {F.}~\bibnamefont
  {Tafti}},\ }\href {\doibase 10.1103/PhysRevB.103.115143} {\bibfield
  {journal} {\bibinfo  {journal} {Physical Review B}\ }\textbf {\bibinfo
  {volume} {103}},\ \bibinfo {pages} {115143} (\bibinfo {year}
  {2021})}\BibitemShut {NoStop}%
\bibitem [{\citenamefont {Ruff}\ \emph {et~al.}()\citenamefont {Ruff},
  \citenamefont {Widmann}, \citenamefont {Lunkenheimer}, \citenamefont
  {Tsurkan}, \citenamefont {Bordács}, \citenamefont {Kézsmárki},\ and\
  \citenamefont {Loidl}}]{ruff_multiferroicity_nodate}%
  \BibitemOpen
  \bibfield  {author} {\bibinfo {author} {\bibfnamefont {E.}~\bibnamefont
  {Ruff}}, \bibinfo {author} {\bibfnamefont {S.}~\bibnamefont {Widmann}},
  \bibinfo {author} {\bibfnamefont {P.}~\bibnamefont {Lunkenheimer}}, \bibinfo
  {author} {\bibfnamefont {V.}~\bibnamefont {Tsurkan}}, \bibinfo {author}
  {\bibfnamefont {S.}~\bibnamefont {Bordács}}, \bibinfo {author}
  {\bibfnamefont {I.}~\bibnamefont {Kézsmárki}}, \ and\ \bibinfo {author}
  {\bibfnamefont {A.}~\bibnamefont {Loidl}},\ }\href {\doibase
  10.1126/sciadv.1500916} {\bibfield  {journal} {\bibinfo  {journal} {Science
  Advances}\ }\textbf {\bibinfo {volume} {1}},\ \bibinfo {pages}
  {e1500916}}\BibitemShut {NoStop}%
\bibitem [{\citenamefont {Fujima}\ \emph {et~al.}(2017)\citenamefont {Fujima},
  \citenamefont {Abe}, \citenamefont {Tokunaga},\ and\ \citenamefont
  {Arima}}]{fujima_thermodynamically_2017}%
  \BibitemOpen
  \bibfield  {author} {\bibinfo {author} {\bibfnamefont {Y.}~\bibnamefont
  {Fujima}}, \bibinfo {author} {\bibfnamefont {N.}~\bibnamefont {Abe}},
  \bibinfo {author} {\bibfnamefont {Y.}~\bibnamefont {Tokunaga}}, \ and\
  \bibinfo {author} {\bibfnamefont {T.}~\bibnamefont {Arima}},\ }\href
  {\doibase 10.1103/PhysRevB.95.180410} {\bibfield  {journal} {\bibinfo
  {journal} {Physical Review B}\ }\textbf {\bibinfo {volume} {95}},\ \bibinfo
  {pages} {180410} (\bibinfo {year} {2017})}\BibitemShut {NoStop}%
\bibitem [{\citenamefont {Bordács}\ \emph {et~al.}(2017)\citenamefont
  {Bordács}, \citenamefont {Butykai}, \citenamefont {Szigeti}, \citenamefont
  {White}, \citenamefont {Cubitt}, \citenamefont {Leonov}, \citenamefont
  {Widmann}, \citenamefont {Ehlers}, \citenamefont {von Nidda}, \citenamefont
  {Tsurkan}, \citenamefont {Loidl},\ and\ \citenamefont
  {Kézsmárki}}]{bordacs_equilibrium_2017}%
  \BibitemOpen
  \bibfield  {author} {\bibinfo {author} {\bibfnamefont {S.}~\bibnamefont
  {Bordács}}, \bibinfo {author} {\bibfnamefont {A.}~\bibnamefont {Butykai}},
  \bibinfo {author} {\bibfnamefont {B.~G.}\ \bibnamefont {Szigeti}}, \bibinfo
  {author} {\bibfnamefont {J.~S.}\ \bibnamefont {White}}, \bibinfo {author}
  {\bibfnamefont {R.}~\bibnamefont {Cubitt}}, \bibinfo {author} {\bibfnamefont
  {A.~O.}\ \bibnamefont {Leonov}}, \bibinfo {author} {\bibfnamefont
  {S.}~\bibnamefont {Widmann}}, \bibinfo {author} {\bibfnamefont
  {D.}~\bibnamefont {Ehlers}}, \bibinfo {author} {\bibfnamefont {H.-A.~K.}\
  \bibnamefont {von Nidda}}, \bibinfo {author} {\bibfnamefont {V.}~\bibnamefont
  {Tsurkan}}, \bibinfo {author} {\bibfnamefont {A.}~\bibnamefont {Loidl}}, \
  and\ \bibinfo {author} {\bibfnamefont {I.}~\bibnamefont {Kézsmárki}},\
  }\href {\doibase 10.1038/s41598-017-07996-x} {\bibfield  {journal} {\bibinfo
  {journal} {Scientific Reports}\ }\textbf {\bibinfo {volume} {7}},\ \bibinfo
  {pages} {7584} (\bibinfo {year} {2017})}\BibitemShut {NoStop}%
\bibitem [{\citenamefont {Kurumaji}\ \emph {et~al.}(2019)\citenamefont
  {Kurumaji}, \citenamefont {Nakajima}, \citenamefont {Hirschberger},
  \citenamefont {Kikkawa}, \citenamefont {Yamasaki}, \citenamefont {Sagayama},
  \citenamefont {Nakao}, \citenamefont {Taguchi}, \citenamefont {Arima},\ and\
  \citenamefont {Tokura}}]{kurumaji_skyrmion_2019}%
  \BibitemOpen
  \bibfield  {author} {\bibinfo {author} {\bibfnamefont {T.}~\bibnamefont
  {Kurumaji}}, \bibinfo {author} {\bibfnamefont {T.}~\bibnamefont {Nakajima}},
  \bibinfo {author} {\bibfnamefont {M.}~\bibnamefont {Hirschberger}}, \bibinfo
  {author} {\bibfnamefont {A.}~\bibnamefont {Kikkawa}}, \bibinfo {author}
  {\bibfnamefont {Y.}~\bibnamefont {Yamasaki}}, \bibinfo {author}
  {\bibfnamefont {H.}~\bibnamefont {Sagayama}}, \bibinfo {author}
  {\bibfnamefont {H.}~\bibnamefont {Nakao}}, \bibinfo {author} {\bibfnamefont
  {Y.}~\bibnamefont {Taguchi}}, \bibinfo {author} {\bibfnamefont {T.-h.}\
  \bibnamefont {Arima}}, \ and\ \bibinfo {author} {\bibfnamefont
  {Y.}~\bibnamefont {Tokura}},\ }\href {\doibase 10.1126/science.aau0968}
  {\bibfield  {journal} {\bibinfo  {journal} {Science}\ }\textbf {\bibinfo
  {volume} {365}},\ \bibinfo {pages} {914} (\bibinfo {year}
  {2019})}\BibitemShut {NoStop}%
\bibitem [{\citenamefont {Leroux}\ \emph {et~al.}(2018)\citenamefont {Leroux},
  \citenamefont {Stolt}, \citenamefont {Jin}, \citenamefont {Pete},
  \citenamefont {Reichhardt},\ and\ \citenamefont
  {Maiorov}}]{leroux_skyrmion_2018}%
  \BibitemOpen
  \bibfield  {author} {\bibinfo {author} {\bibfnamefont {M.}~\bibnamefont
  {Leroux}}, \bibinfo {author} {\bibfnamefont {M.~J.}\ \bibnamefont {Stolt}},
  \bibinfo {author} {\bibfnamefont {S.}~\bibnamefont {Jin}}, \bibinfo {author}
  {\bibfnamefont {D.~V.}\ \bibnamefont {Pete}}, \bibinfo {author}
  {\bibfnamefont {C.}~\bibnamefont {Reichhardt}}, \ and\ \bibinfo {author}
  {\bibfnamefont {B.}~\bibnamefont {Maiorov}},\ }\href {\doibase
  10.1038/s41598-018-33560-2} {\bibfield  {journal} {\bibinfo  {journal}
  {Scientific Reports}\ }\textbf {\bibinfo {volume} {8}},\ \bibinfo {pages}
  {15510} (\bibinfo {year} {2018})}\BibitemShut {NoStop}%
\bibitem [{\citenamefont {Lin}\ \emph {et~al.}(2016)\citenamefont {Lin},
  \citenamefont {Li}, \citenamefont {Deng}, \citenamefont {Xing}, \citenamefont
  {Liu}, \citenamefont {Zhu}, \citenamefont {Yang},\ and\ \citenamefont
  {Wen}}]{lin_multiband_2016}%
  \BibitemOpen
  \bibfield  {author} {\bibinfo {author} {\bibfnamefont {H.}~\bibnamefont
  {Lin}}, \bibinfo {author} {\bibfnamefont {Y.}~\bibnamefont {Li}}, \bibinfo
  {author} {\bibfnamefont {Q.}~\bibnamefont {Deng}}, \bibinfo {author}
  {\bibfnamefont {J.}~\bibnamefont {Xing}}, \bibinfo {author} {\bibfnamefont
  {J.}~\bibnamefont {Liu}}, \bibinfo {author} {\bibfnamefont {X.}~\bibnamefont
  {Zhu}}, \bibinfo {author} {\bibfnamefont {H.}~\bibnamefont {Yang}}, \ and\
  \bibinfo {author} {\bibfnamefont {H.-H.}\ \bibnamefont {Wen}},\ }\href
  {\doibase 10.1103/PhysRevB.93.144505} {\bibfield  {journal} {\bibinfo
  {journal} {Physical Review B}\ }\textbf {\bibinfo {volume} {93}},\ \bibinfo
  {pages} {144505} (\bibinfo {year} {2016})}\BibitemShut {NoStop}%
\bibitem [{\citenamefont {Nikolić}(2020)}]{nikolic_quantum_2020}%
  \BibitemOpen
  \bibfield  {author} {\bibinfo {author} {\bibfnamefont {P.}~\bibnamefont
  {Nikolić}},\ }\href {\doibase 10.1103/PhysRevB.102.075131} {\bibfield
  {journal} {\bibinfo  {journal} {Physical Review B}\ }\textbf {\bibinfo
  {volume} {102}},\ \bibinfo {pages} {075131} (\bibinfo {year}
  {2020})}\BibitemShut {NoStop}%
\bibitem [{\citenamefont {Rourke}\ and\ \citenamefont
  {Julian}(2012)}]{rourke_numerical_2012}%
  \BibitemOpen
  \bibfield  {author} {\bibinfo {author} {\bibfnamefont {P.~M.~C.}\
  \bibnamefont {Rourke}}\ and\ \bibinfo {author} {\bibfnamefont {S.~R.}\
  \bibnamefont {Julian}},\ }\href {\doibase 10.1016/j.cpc.2011.10.015}
  {\bibfield  {journal} {\bibinfo  {journal} {Computer Physics Communications}\
  }\textbf {\bibinfo {volume} {183}},\ \bibinfo {pages} {324} (\bibinfo {year}
  {2012})}\BibitemShut {NoStop}%
\bibitem [{\citenamefont {Yang}\ \emph {et~al.}(2020)\citenamefont {Yang},
  \citenamefont {Singh}, \citenamefont {Lu}, \citenamefont {Huang},
  \citenamefont {Bahrami}, \citenamefont {Chiu}, \citenamefont {Graf},
  \citenamefont {Huang}, \citenamefont {Wang}, \citenamefont {Lin},
  \citenamefont {Torchinsky}, \citenamefont {Bansil},\ and\ \citenamefont
  {Tafti}}]{yang_transition_2020}%
  \BibitemOpen
  \bibfield  {author} {\bibinfo {author} {\bibfnamefont {H.-Y.}\ \bibnamefont
  {Yang}}, \bibinfo {author} {\bibfnamefont {B.}~\bibnamefont {Singh}},
  \bibinfo {author} {\bibfnamefont {B.}~\bibnamefont {Lu}}, \bibinfo {author}
  {\bibfnamefont {C.-Y.}\ \bibnamefont {Huang}}, \bibinfo {author}
  {\bibfnamefont {F.}~\bibnamefont {Bahrami}}, \bibinfo {author} {\bibfnamefont
  {W.-C.}\ \bibnamefont {Chiu}}, \bibinfo {author} {\bibfnamefont
  {D.}~\bibnamefont {Graf}}, \bibinfo {author} {\bibfnamefont {S.-M.}\
  \bibnamefont {Huang}}, \bibinfo {author} {\bibfnamefont {B.}~\bibnamefont
  {Wang}}, \bibinfo {author} {\bibfnamefont {H.}~\bibnamefont {Lin}}, \bibinfo
  {author} {\bibfnamefont {D.}~\bibnamefont {Torchinsky}}, \bibinfo {author}
  {\bibfnamefont {A.}~\bibnamefont {Bansil}}, \ and\ \bibinfo {author}
  {\bibfnamefont {F.}~\bibnamefont {Tafti}},\ }\href {\doibase
  10.1063/1.5132958} {\bibfield  {journal} {\bibinfo  {journal} {APL
  Materials}\ }\textbf {\bibinfo {volume} {8}},\ \bibinfo {pages} {011111}
  (\bibinfo {year} {2020})}\BibitemShut {NoStop}%
\end{thebibliography}%

\end{document}